\documentclass[3p,times]{elsarticle}

\usepackage{ecrc}

\newcommand{\be}{\begin{equation}}
\newcommand{\ee}{\end{equation}}
\newcommand{\bea}{\begin{eqnarray}}
\newcommand{\eea}{\end{eqnarray}}
\newcommand{\mybibitem}{\bibitem}
\newcommand{\gton}{\mathrel{\lower.9ex \hbox{$\stackrel{\displaystyle 
>}{\sim}$}}} 
\newcommand{\lton}{\mathrel{\lower.9ex \hbox{$\stackrel{\displaystyle 
<}{\sim}$}}}  

\volume{00}

\firstpage{1}

\journalname{Nuclear Physics A}

\runauth{}

\jid{npa}

\jnltitlelogo{Nuclear Physics A}

\usepackage{amssymb}
\biboptions{square,comma,numbers,sort&compress}

\usepackage[figuresright]{rotating}

\begin{document}

\begin{frontmatter}

%% Title, authors and addresses

%% use the tnoteref command within \title for footnotes;
%% use the tnotetext command for the associated footnote;
%% use the fnref command within \author or \address for footnotes;
%% use the fntext command for the associated footnote;
%% use the corref command within \author for corresponding author footnotes;
%% use the cortext command for the associated footnote;
%% use the ead command for the email address,
%% and the form \ead[url] for the home page:
%%
%% \title{Title\tnoteref{label1}}
%% \tnotetext[label1]{}
%% \author{Name\corref{cor1}\fnref{label2}}
%% \ead{email address}
%% \ead[url]{home page}
%% \fntext[label2]{}
%% \cortext[cor1]{}
%% \address{Address\fnref{label3}}
%% \fntext[label3]{}

\dochead{}
%% Use \dochead if there is an article header, e.g. \dochead{Short communication}
%% \dochead can also be used to include a conference title, if directed by the editors
%% e.g. \dochead{17th International Conference on Dynamical Processes in Excited States of Solids}

\title{Interplay between bulk medium evolution and (D)GLV energy loss}

%% use optional labels to link authors explicitly to addresses:
%% \author[label1,label2]{<author name>}
%% \address[label1]{<address>}
%% \address[label2]{<address>}

\author{Denes Molnar and Deke Sun}

\address{Department of Physics and Astronomy, Purdue University, 
West Lafayette, IN 47907, U.S.A. }

\begin{abstract}
We study the consistency between high-$p_T$ 
nuclear suppression ($R_{AA}$) and elliptic
flow ($v_2$) using Gyulassy-Levai-Vitev (GLV)
energy loss or a simpler power-law $dE/dL$ formula,
for a variety of bulk evolution models.
The results generally confirm our earlier work~\cite{GLV3d} that found
suppressed elliptic flow for transversely expanding media.
One exception is the set of hydrodynamic solutions used recently\cite{BetzGyulassy} 
by Betz and Gyulassy, which give significantly higher $v_2$ but unfortunately
assume unrealistic bag-model equation of state. On the other hand, 
we show that 
covariant treatment of energy loss introduces an interplay 
between jet direction and hydrodynamic flow of the medium, which largely 
counteracts elliptic flow suppression caused by transverse expansion.
\end{abstract}

\begin{keyword}
Heavy-ion collisions, elliptic flow, parton energy loss
%% keywords here, in the form: keyword \sep keyword

%% PACS codes here, in the form: \PACS code \sep code

%% MSC codes here, in the form: \MSC code \sep code
%% or \MSC[2008] code \sep code (2000 is the default)

\end{keyword}

\end{frontmatter}

%%
%% Start line numbering here if you want
%%
% \linenumbers

%% main text
\section{Introduction}
\label{Sc:intro}

An important crosscheck of parton energy loss calculations is the consistency
between nuclear suppression ($R_{AA}$) and differential elliptic flow 
$v_2(p_T)$.
Recently we found\cite{GLV3d} 
that in realistic applications of 
Gyulassy-Levai-Vitev (GLV) radiative parton energy loss\cite{GLV}
that include transverse expansion of the bulk 
medium, high-$p_T$ elliptic flow is reduced by nearly a half 
compared to transversely frozen evolution scenarios.
This reinforced the conclusions\cite{PHENIXeloss} by the PHENIX Collaboration
that perturbative QCD energy loss models generally 
fail to reproduce the azimuthal angle
dependent neutral pion suppression.
However, a recent work by Betz and Gyulassy claims\cite{BetzGyulassy} 
simultaneous reproduction of this set of observables with simple 
pQCD-motivated energy loss formulas. 
This apparent contradiction, on the other hand,
may be due to important differences between the two calculations,
especially in the energy loss model and bulk medium evolution assumed.
Here we pinpoint
the origin of the discrepancy, and 
show that the findings of Ref.~\cite{BetzGyulassy} are largely due to
the hydrodynamic solutions
used in that calculation for bulk medium evolution. In addition, we show that 
covariant treatment of energy loss introduces an interplay 
between jet direction and hydrodynamic flow of the medium, which largely 
compensates the elliptic flow suppression
we found earlier in~\cite{GLV3d}.

%%%%
\section{Radiative energy loss and bulk medium evolution}
\label{Sc:rad}

%Technically Ref.~\cite{GLV3d} calculated  $v_2(p_T)$ and
%azimuth-averaged $R_{AA}(p_T)$,
%while Ref.~\cite{BetzGyulassy} computed $R_{AA}(\phi)$, but that is fine
%because we confirmed that the PHENIX data sets
%for these observables are consistent with each other if one assumes pure
%elliptic flow azimuthal dependence,
%$dN/d\phi \propto 1 + 2v_2(p_T) \cos 2\phi$. 
%%
\subsection{Sensitivity to bulk medium model}

Consider the parameterized
energy loss model $dE/dL = \kappa \, E^a L^b T^c$
by Betz and Gyulassy\cite{BetzGyulassy},
with ``pQCD-like'' exponents $a = 1/3$, $b = 1$, 
and $c = 2 - a + b = 8/3$ ($\kappa$ is then dimensionless). 
Here $E$ is the jet energy, 
$T$ is temperature of the medium, 
and $L$ is the pathlength traveled by the jet.
To study the sensitivity to the bulk medium evolution,
we investigate five different dynamical models for Au+Au at $\sqrt{s_{NN}} = 200$~GeV at RHIC, impact parameter $b \approx 7.5$~fm.
Four of these are solutions of boost-invariant 2+1D hydrodynamics 
using the VISH2+1 code\cite{VISH2p1}, which are available in
tabulated form from the TECHQM Collaboration website~\cite{TECHQM} in two data 
sets.
Set 1 is for a ``bag-model'' like equation of state (EoS), 
``fKLN'' initial profile motivated by the color glass condensate model, 
with zero viscosity or constant 
$\eta/s = 0.08$. The ideal and viscous versions of this set
are practically identical for observables studied here,
so we only show results for ``ideal-fKLN'',
which is the evolution used in Ref.~\cite{BetzGyulassy}.
Set 2 from TECHQM is a later calculation with more 
realistic lattice QCD EoS, constant $\eta/s = 0.08$,
for fKLN or Glauber initial profile.
The fifth model is the same covariant transport evolution as
in Ref.~\cite{GLV3d}, computed using 
Molnar's Parton Cascade (MPC) \cite{MPC}.

\begin{figure}[t]
\leavevmode
\begin{center}
\includegraphics[height=50mm]{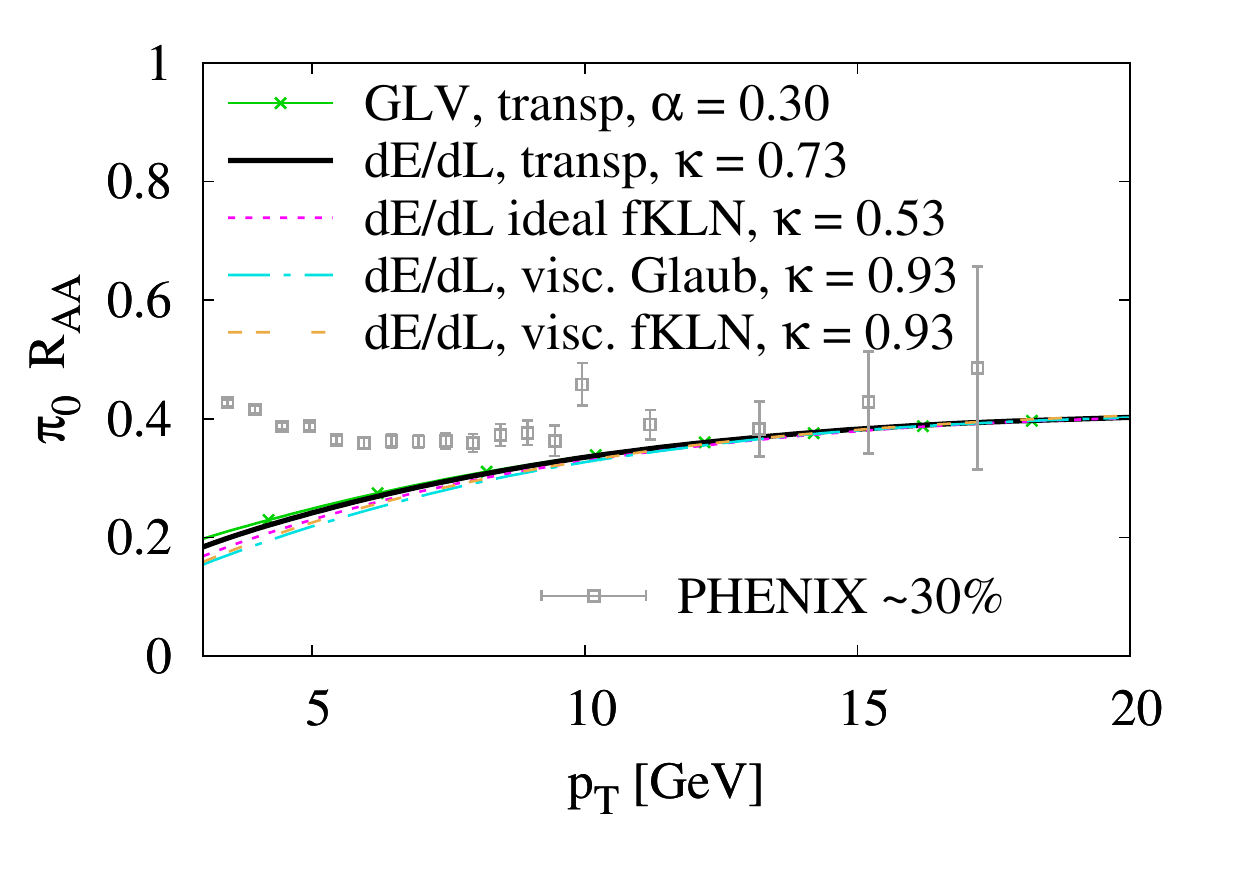}
\hskip 0cm
\includegraphics[height=50mm]{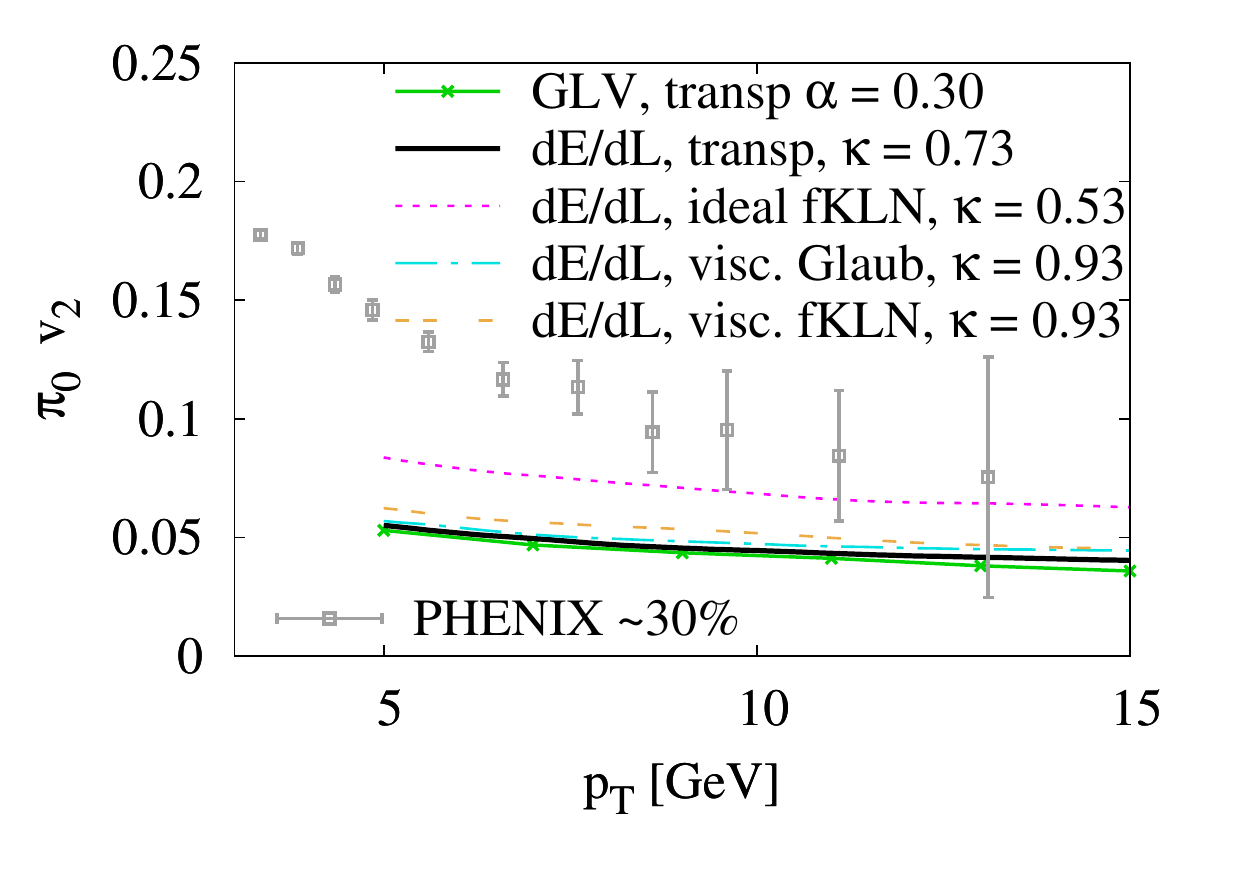}
\end{center}
\caption{Neutral pion suppression factor $R_{AA}$ (left) and differential
elliptic flow $v_2(p_T)$ (right) 
at midrapidity in mid-peripheral ($b\approx 7.5 fm$)
Au+Au at $\sqrt{s_{NN}} = 200$ GeV, calculated using parton energy loss 
parameterization\cite{BetzGyulassy} 
$dE/dL = \kappa E^{1/3} L^1 T^{8/3}$. Results for four different 
bulk medium models are plotted (see text):
i) ideal hydrodynamics with fKLN initial profile from ``Set 1'' (dotted);
ii) viscous hydrodynamics with $\eta/s = 0.08$ and 
fKLN initial profile (dashed-dotted);
iii) viscous hydrodynamics with $\eta/s = 0.08$ and 
Glauber initial profile from ``Set 2'' (double short dashes);
and iv) covariant parton transport MPC as in Ref.~\cite{GLV3d} 
(solid lines).
For comparison, results from Ref.~\cite{GLV3d} using MPC and GLV energy loss 
are also shown (solid lines with crosses). 
In all cases energy loss is scaled to set a fixed $R_{AA} \sim 0.4$ at $p_T \sim 15-20$~GeV. 
As in Ref.~\cite{GLV3d},
data\cite{PHENIX_pi0_y2008,PHENIX_pi0_v2_y2010} from PHENIX (boxes) are shown to guide the eye.
}
\label{fig:1}
\end{figure}
Figure~\ref{fig:1} shows the neutral pion $R_{AA}$ (left plot) and
$v_2$ (right plot) in $Au+Au$ at RHIC for these scenarios. 
$R_{AA}$ is basically
the same for all cases because $\kappa$ is
dialed to obtain the same suppression at high $p_T$. 
In all scenarios, elliptic flow is reduced to $\sim 4-5$\% at high $p_T$,
much the same value as what we found earlier with GLV energy loss\cite{GLV3d},
{\em except} for the ``ideal-fKLN'' evolution studied by Betz and Gyulassy.
Thus we confirm their result, but also demonstrate that transverse
expansion does suppress $v_2$ for a hydrodynamic medium as well if one includes
a realistic equation of state.

\subsection{Energy loss model}

Next we test how well the power-law $dE/dL \propto E^a L^b T^c$ formula 
captures perturbative QCD parton energy loss in the Gyulassy-Levai-Vitev (GLV) 
formulation\cite{GLV}. 
The approach is identical to the one in Ref.~\cite{GLV3d}, i.e.,
we use the average radiative
energy loss along the path of a massless jet parton
obtained via integrating
the first-order (in opacity) GLV radiated gluon spectrum:
\be
\langle \Delta E^{(1)}\rangle 
=  \frac{C_R \alpha_s}{\pi^2} E
\int\limits_0^\infty  d\tau \rho({\vec x}_0 + {\vec v}\tau, \tau)\sigma_{gg}(\tau)
 \int dx\, d^2{\bf k}\,
\int d^2 {\bf q} \, 
\frac{\mu^2(\tau)}{\pi [{\bf q}^2 + \mu^2(\tau)]^2}
\, \frac{2 {\bf k}{\bf q}}{{\bf k}^2 ({\bf k} - {\bf q})^2}
\left( 1 -   \cos \frac{({\bf k} - {\bf q})^2 \tau}{2xE} \right) 
\ ,
\label{GLV1}
\ee
where $E$ is the jet
parton energy, $\bf q$ is the momentum transfer in scattering with the
medium, $\mu$ is the local Debye screening mass, 
$\sigma_{gg}=9\pi\alpha_s^2/2\mu^2$ is the scattering cross section in 
the medium for gluons,
and the momentum integrals are performed
observing finite energy and kinematic bounds 
($|k|\  \lton x E$, $|q| \ \lton  \sqrt{6 E T}$, $x E \ \gton \mu$).

\begin{figure}[t]
\leavevmode
\begin{center}
\includegraphics[height=50mm]{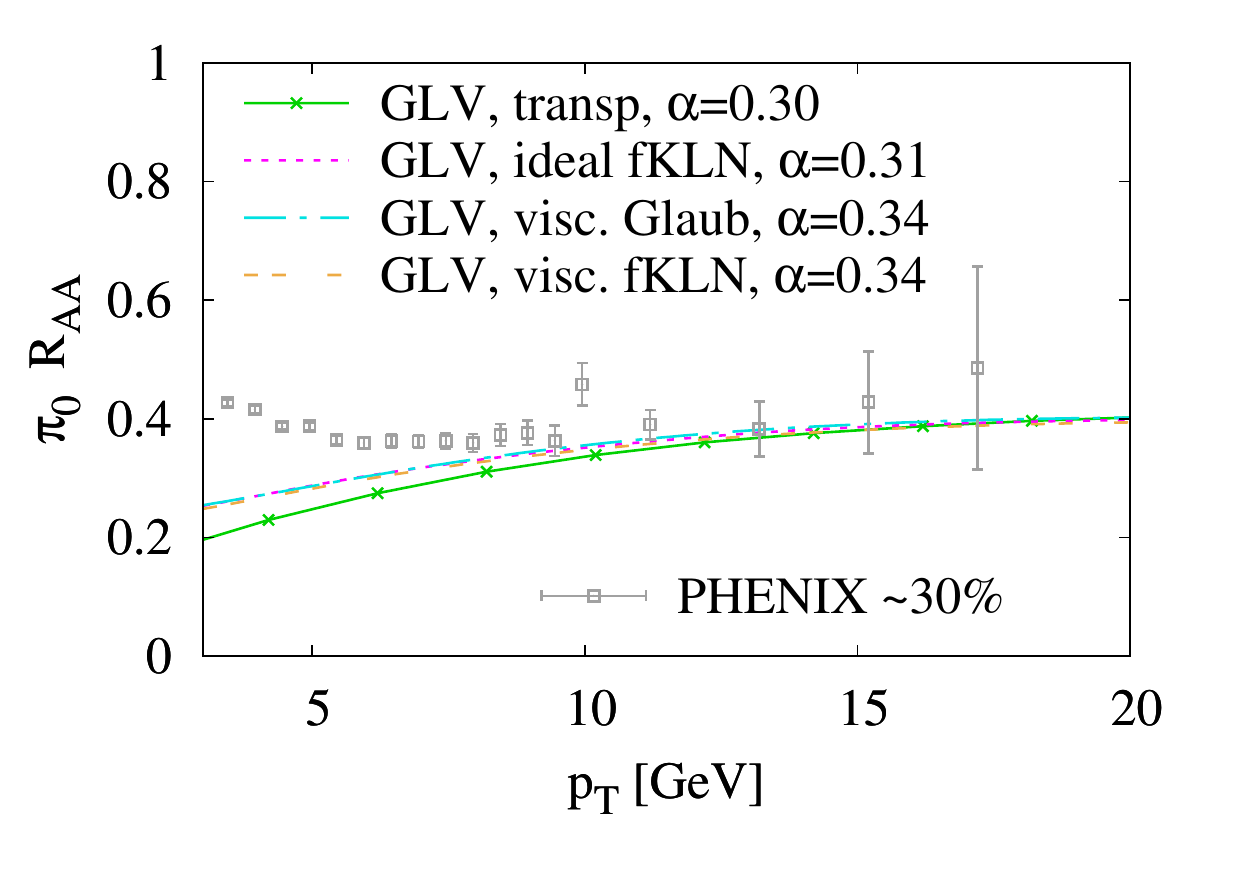}
\hskip 0cm
\includegraphics[height=50mm]{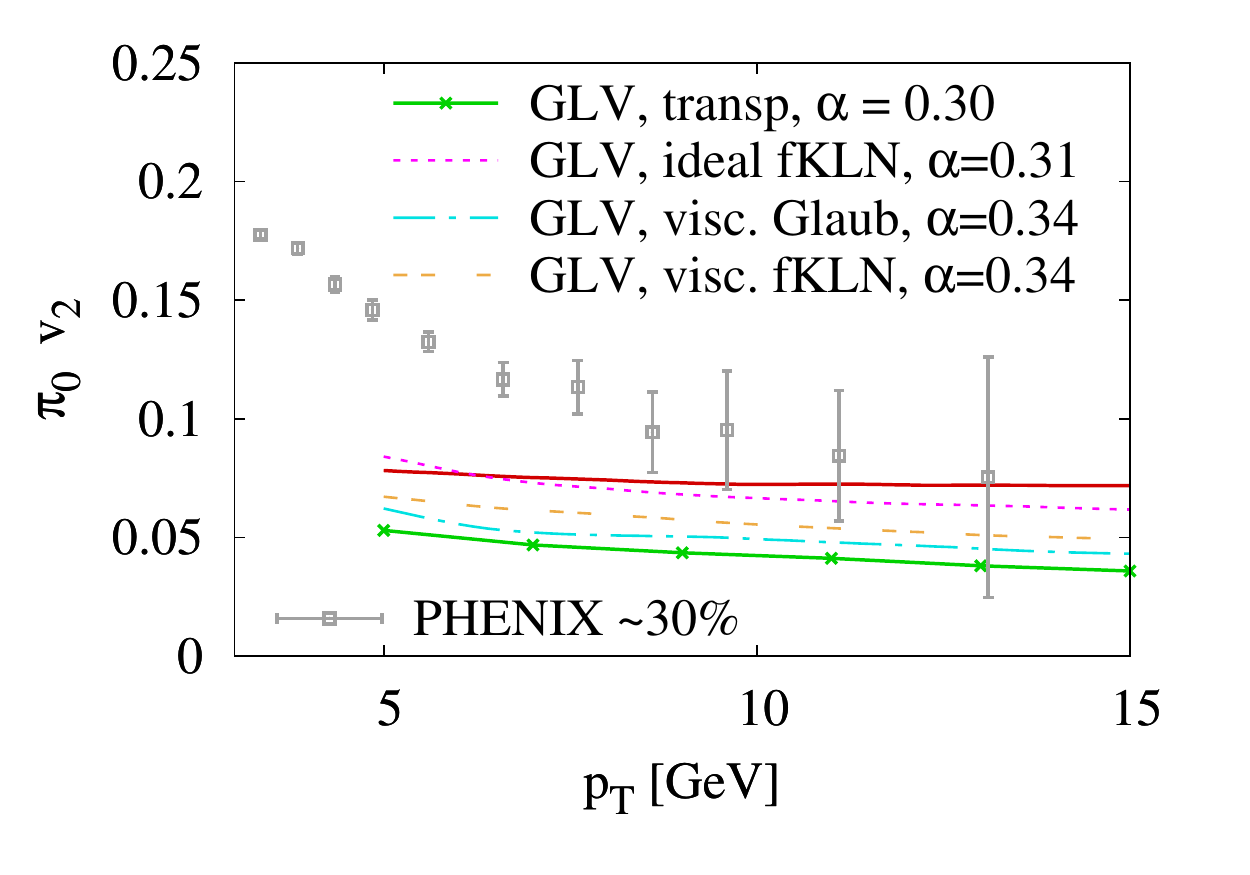}
\end{center}
\caption{The same as Fig.~\ref{fig:1}, except calculated using GLV energy loss.The solid line (without crosses) in the right plot 
now shows $v_2$ from Ref.~\cite{GLV3d} for transversely 
frozen, boost-invariant 0+1D medium evolution. }
\label{fig:2}
\end{figure}
Figure~\ref{fig:2} shows neutral pion $R_{AA}$ and $v_2$ for the different
bulk medium scenarios with GLV energy loss. Qualitatively the results
are very similar to those in Fig.~\ref{fig:1}, confirming that the ``pQCD-like'' exponents in Ref.~\cite{BetzGyulassy} are a 
reasonable approximation to GLV energy loss. 
After fixing $R_{AA} \sim 0.4$
at high $p_T$ (left plot), a residual sensitivity to the bulk
evolution still remains in elliptic
flow (right plot). 
The ``ideal-fKLN'' evolution used in Ref.~\cite{BetzGyulassy} gives
largest $v_2$, almost as large as the results 
with transversely frozen dynamics in Ref.~\cite{GLV3d}
(solid line). Hydrodynamic solutions with lattice QCD EoS, on the other hand,
give smaller $v_2$. There is a modest $\sim 15$\% difference
between fKLN and Glauber profiles with viscous hydrodynamics (fKLN is higher),
which may help 
constrain the initial geometry.

\subsection{Covariant energy loss}

Neither of the above calculations observe proper Lorentz covariance, however,
because both $dE/dL \propto E^a L^b T^c$ and GLV energy loss 
Eq.~(\ref{GLV1}) give frame
dependent results. We can formulate a frame-independent prescription
if we require energy loss contributions 
to be computed in the frame where the fluid is locally static 
along the path (LR frame). For massless partons produced at spacetime point
$(0,\vec 0)$, scattering occurs at $L (1, \vec v)$, which transforms the same
way as the four-momentum $E(1,\vec v)$. Therefore, in the massless case 
$dE/dL$ is a Lorentz scalar,
which means that for the $dE/dL$ model we should have
\be
\frac{dE}{dL} = \frac{dE_{LR}}{dL_{LR}} 
= \kappa\, E_{LR}^a\, L_{LR}^b T^c = 
\kappa\, [\gamma_F ( 1 - \vec v\vec v_F)]^{a+b}\, E^a L^b T^c \ ,
\label{cov_dEdL}
\ee
while for GLV
\be
dL_{LR}\, \rho_{LR} \, \sigma =
dL\, \rho_{LR}\, \sigma\, \gamma_F (1 - {\vec v} {\vec v_F})
= dL\, \rho\, \sigma \, (1 - {\vec v} {\vec v_F}) \ .
\label{cov_opacity}
\ee
Here, ${\vec v}_F$ is the local three-velocity of fluid flow, while 
$\gamma_F \equiv (1- v_F^2)^{-1/2}$. In both cases, a new factor appears
that couples the motion of the jet to that of the fluid.
For GLV this is very similar to the term introduced in Ref.~\cite{BaierFlow},
however, in contrast to the results there 
we find that jet-medium flow coupling has significant effect on observables.

\begin{figure}[t]
\leavevmode
\begin{center}
\includegraphics[height=50mm]{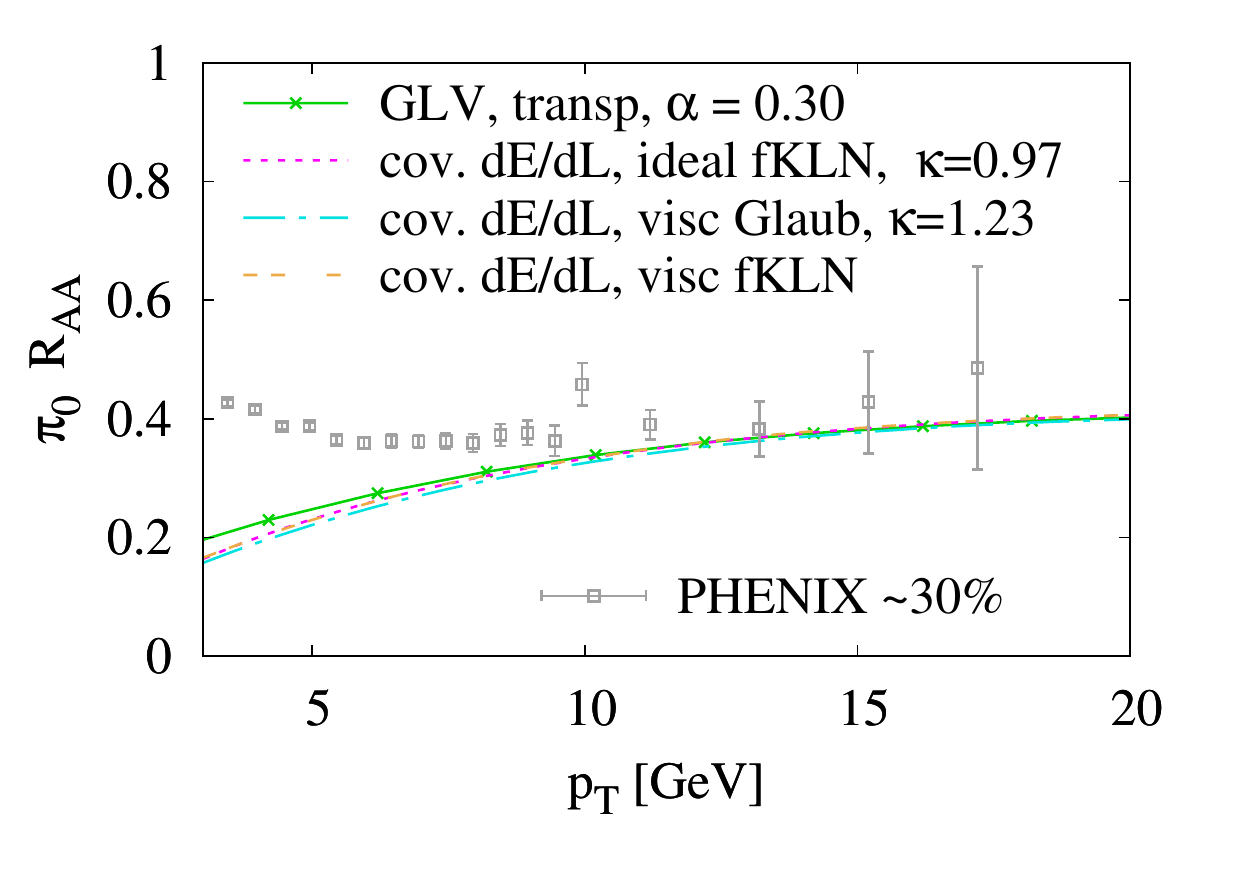}
\hskip 0cm
\includegraphics[height=50mm]{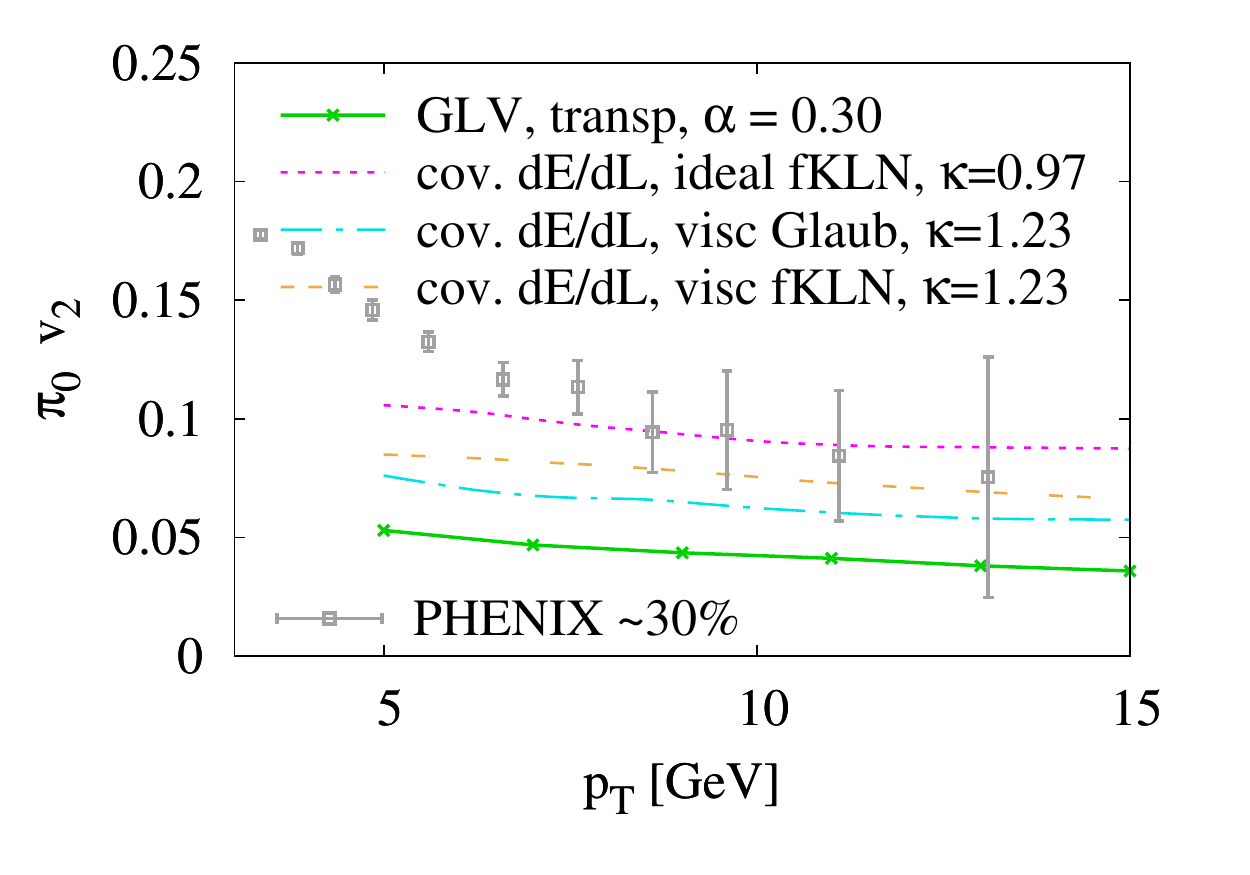}
\end{center}
\caption{The same as Fig.~\ref{fig:1}, except for a covariant $dE/dL$
calculation using Eq. (\ref{cov_dEdL}).}
\label{fig:3}
\end{figure}
Figure~\ref{fig:3} shows neutral pion $R_{AA}$ and $v_2$ in Au+Au at RHIC
with $b\approx 7.5$~fm, calculated using covariant $dE/dL$ energy loss. Two
features are noticeable immediately. First, with covariant energy
loss one needs higher scaling factors $\kappa$ to obtain the same $R_{AA}$.
Second, even after setting $\kappa$ to $R_{AA}$ at high $p_T$, 
$v_2$ is larger with covariant energy loss and shows strong
dependence on bulk dynamics.
Qualitatively the reason is that
jet-medium flow coupling reduces energy loss for jets moving
in the same direction as the medium flow, 
the more the larger the flow velocity. 
For jets that move in-plane (short direction), flow tends to be larger,
so the reduction is stronger.
The resulting $v_2$ enhancement largely cancels out the
flow suppression due to transverse expansion found in Ref.~\cite{GLV3d}.
We find the largest $v_2$ for the ``ideal-fKLN'' profile used in 
Ref.~\cite{BetzGyulassy}.
 
\begin{figure}[t]
\leavevmode
\begin{center}
\includegraphics[height=50mm]{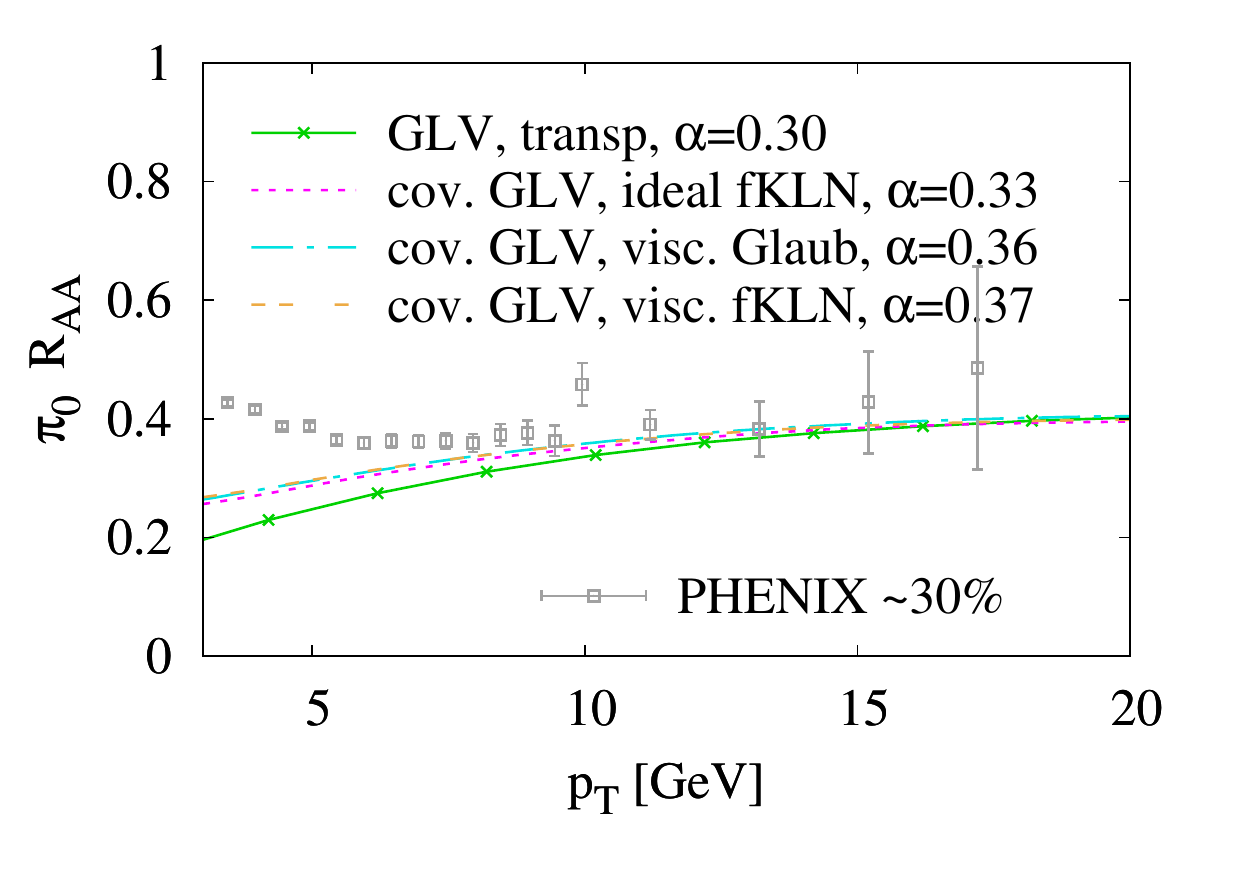}
\hskip 0cm
\includegraphics[height=50mm]{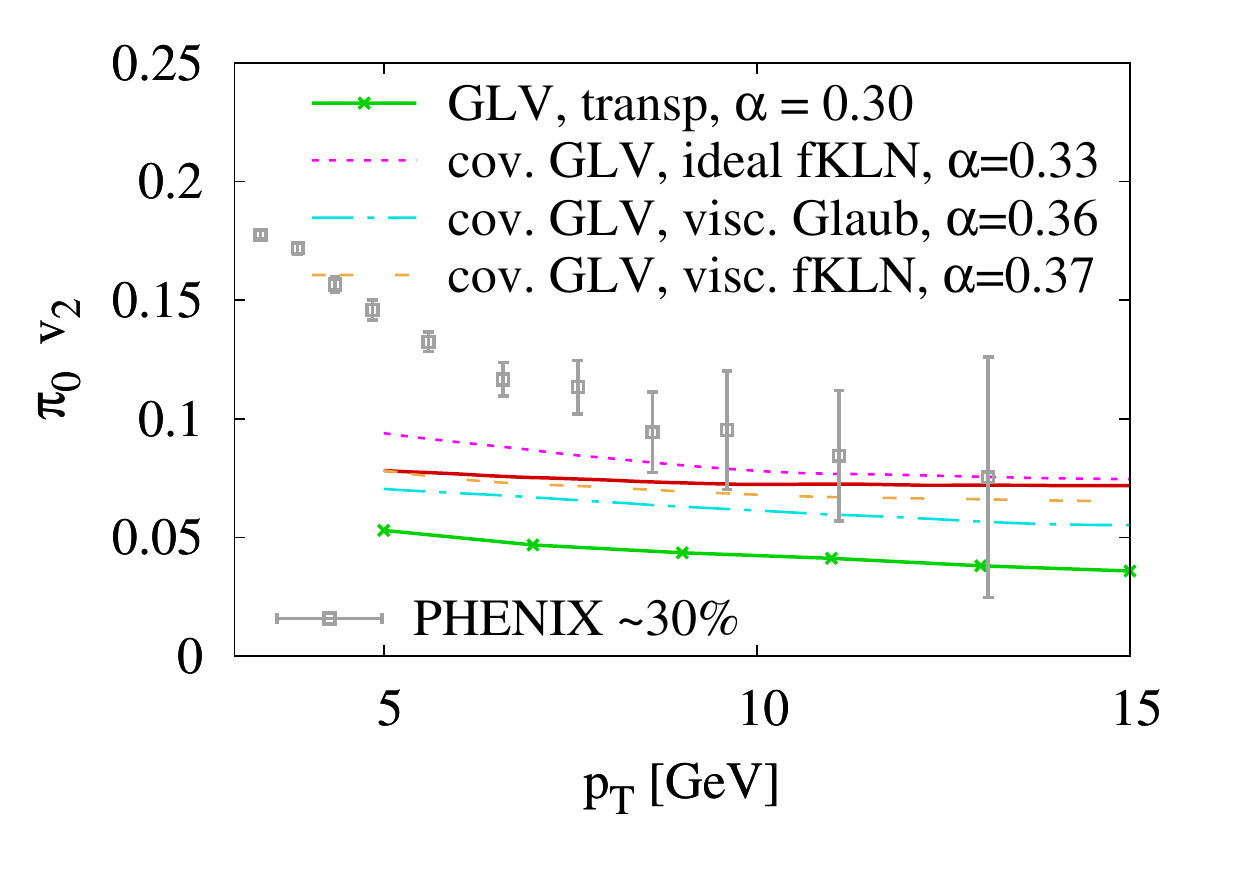}
\end{center}
\caption{The same as Fig.~\ref{fig:2}, except with the covariant opacity
factor Eq. (\ref{cov_dEdL}) in the GLV energy loss formula Eq.~(\ref{GLV1}).}
\label{fig:4}
\end{figure}
Very similar results follow with covariant GLV energy loss,
as shown in
Figure~\ref{fig:4}. Elliptic flow is a little bit smaller than
for the covariant $dE/dL$ model but otherwise it
shows the same ordering between the various scenarios.

At the conference we also showed preliminary results for charm and bottom 
quarks
with Djordjevic-Gyulassy-Levai-Vitev (DGLV) energy loss\cite{DGLV}. 
Due to space constraints these results will be presented elsewhere.

%%%%
\section{Conclusions}
\label{Sc:conc}

We study the consistency between high-$p_T$ 
nuclear suppression ($R_{AA}$) and elliptic
flow ($v_2$) using Gyulassy-Levay-Vitev 
energy loss or a simpler power-law $dE/dL$ formula,
for a variety of bulk evolution models.
The results generally confirm our earlier work~\cite{GLV3d} that found
suppressed elliptic flow for transversely expanding media.
However,
one exception is the set of hydrodynamic solutions used recently\cite{BetzGyulassy} 
by Betz and Gyulassy, which give significantly higher $v_2$ but unfortunately
assume unrealistic bag-model equation of state. On the other hand, 
we also find that 
covariant treatment of energy loss introduces an interplay 
between jet direction and hydrodynamic flow of the medium, which largely 
compensates for the elliptic flow suppression
we found earlier in~\cite{GLV3d}.

%% The Appendices part is started with the command \appendix;
%% appendix sections are then done as normal sections
%% \appendix

%% \section{}
%% \label{}

%% References
%%
%% Following citation commands can be used in the body text:
%% Usage of \cite is as follows:
%%   \cite{key}         ==>>  [#]
%%   \cite[chap. 2]{key} ==>> [#, chap. 2]
%%

%% References with BibTeX database:

%\bibliographystyle{elsarticle-num}
%\bibliography{<your-bib-database>}

%% Authors are advised to use a BibTeX database file for their reference list.
%% The provided style file elsarticle-num.bst formats references in the required Procedia style

%% For references without a BibTeX database:

{\bf Acknowledgements:}
This work was supported by the US DOE under grant
DE-PS02-09ER41665. D.S. was partially supported by the JET Collaboration
(DOE grant DE-AC02-05CH11231).

\end{document}